\documentclass[a4paper,11pt]{article}
\usepackage{pos}
\usepackage{setspace}
\usepackage{wrapfig}
\usepackage[utf8]{inputenc}

\newcommand{\Xmax}{X_\text{max}}

\newcommand{\myparagraph}[1]{\vspace{5pt} \noindent \textbf{#1}}

\renewenvironment{thebibliography}[1]{
\begin{oldthebibliography}{#1}
  \setlength{\itemsep}{0em}
  \setlength{\parskip}{0em}
}
{
  \end{oldthebibliography}
}

\title{Science with the Global Cosmic-ray Observatory (GCOS)}
\ShortTitle{Science with GCOS}

\manuallySeparateAuthors
\author*[a,b]{Rafael~{Alves Batista}}
\author{ for the GCOS Collaborators}

\affiliation[a]{Instituto de Física Teórica UAM-CSIC\\ C/ Nicolás Cabrera 13-15, 28049, Madrid, Spain}
\affiliation[b]{Departamento de Física Teórica, Universidad Autónoma de Madrid\\
M-15, 28049, Madrid, Spain}

\emailAdd{rafael.alvesbatista@uam.es}

\abstract{
The Global Cosmic-ray Observatory (GCOS) is a proposed large-scale observatory for studying ultra-high-energy cosmic particles, including ultra-high-energy cosmic rays (UHECRs), photons, and neutrinos. Its primary goal is to characterise the properties of the highest-energy particles in Nature with unprecedented accuracy, and to identify their elusive sources. With an aperture at least a ten-fold larger than existing observatories, this next-generation facility should start operating after 2030, when present-day detectors will gradually cease their activities. Here we briefly review the scientific case motivating GCOS. We present the status of the project, preliminary ideas for its design, and some estimates of its capabilities. 
}

\ConferenceLogo{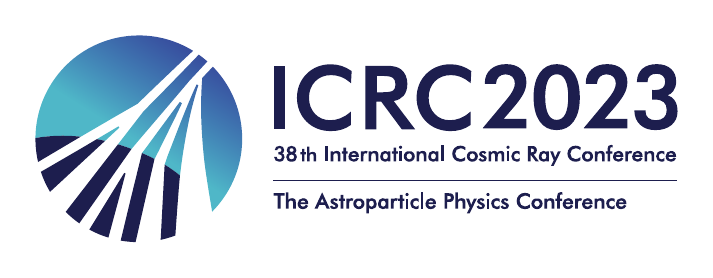}

\FullConference{%
The 38th International Cosmic Ray Conference (ICRC2023)\\
  26 July -- 3 August, 2023\\
  Nagoya, Japan}

\begin{document}
\maketitle

\section{Introduction}
\label{sec:intro}

\vspace{-0.1cm}

Since the first detection of particles with energies exceeding 1~EeV over half a century ago~\cite{linsley1961a, linsley1963a}, considerable resources have been invested in trying to understand the origins of the most energetic particles known in nature, the ultra-high-energy cosmic rays (UHECRs). At energies $E \gtrsim 60 \; \text{EeV}$, their flux is as low as $1 \; \text{particle} \, \text{km}^{-2} \, \text{century}^{-1}$~\cite{alvesbatista2019d, anchordoqui2019a}. Yet, pioneering experiments succeeded in instrumenting areas of up to hundreds of $\text{m}^2$ to collect these elusive particles~\cite{clark1957a, linsley1961a}. Present-day experiments such as the Telescope Array (TA) and the Pierre Auger Observatory have managed to deploy their detectors over areas of $700 \; \text{km}^2$ (soon to be $2800 \; \text{km}^2$) and $3000 \; \text{km}^2$, respectively~\cite{coleman2023a}.

Significant progress has been made with these observatories, but the full picture is still blurry. \emph{Where are UHECRs produced?} \emph{How are they accelerated to ultra-high energies?} These questions have been partly driving the astrophysical interest in this field. By-products of UHECR interactions with gas and radiation, whether in their sources or in the intergalactic medium, provide vital pieces of information to make sense of high-energy observations, particularly of gamma rays and neutrinos, thereby establishing a link between them. Notably, hadronic processes are generally necessary to generate these connections, with CRs playing an essential role. Evidently, signatures at lower energies and potential correlations with gravitational waves are possible. This complementarity highlights the integral role of UHECRs within the multimessenger paradigm, as they are crucial for interpreting multimessenger observations at high energies.

\emph{Which particles contribute to the UHECR flux?} On one hand, this question poses difficulties for astrophysical studies because the UHECR composition is not precisely known on an event-by-event basis. The observable commonly used to infer the UHECR composition, the depth of the shower maximum ($\Xmax$), is only known statistically. However, this question also harks back to the roots of cosmic-ray research in particle physics. The extensive air showers (EASs) triggered by UHECRs interacting with the atmosphere serve as a natural laboratory, surpassing by several orders of magnitude the maximal energy attainable by the Large Hadron Collider (LHC), placing UHECR observatories in a unique position for studies such as the measurement of the cross section of proton interacting with air at high energies~\cite{auger2012d, ta2020b}.



All the studies mentioned above are ongoing using data of current facilities. Auger is undergoing improvements towards AugerPrime~\cite{castellina2019a}, while TA is deploying more detectors to increase its area by a factor 4 (TAx4)~\cite{kido2022a}. However, the future of this field remains uncertain as presently there are no concrete plans to extend operations beyond 2035. For this reason, in 2021, the first workshop of the Global Cosmic-ray Observatory (GCOS)~\cite{hoerandel2022a} was held online. Since then, two other workshops have been organised to gather ideas on how to design a next-generation UHECR observatory capable of addressing the open challenges in this field. Here, we present some preliminary ideas for GCOS, its design, and its scientific goals. More details can be found in the compendium of ideas we compiled following the discussions from the workshops~\cite{gcosCompendium}.

\section{Science Case}
\label{sec:science}

\myparagraph{UHE multimessenger astrophysics.} GCOS' primary goal is to study astrophysics at ultra-high energies, revealing the sources of the most energetic particles in the universe and their inner workings. Nevertheless, it will be a true multimessenger facility, detecting CRs, photons, and neutrinos, despite being optimised for the former. 
GCOS' large exposure will improve considerably the significance of today's results. By simply extrapolating the event rate, the correlations with Centaurus~A and starburst galaxies hinted at by the Auger data~\cite{auger2022d} would reach a $5 \sigma$ level within 1.5~years if the total area exceeds our benchmark of $60,000 \; \text{km}^2$. Scaling arguments suggest that the possible correlation with the Perseus-Pisces supercluster reported by TA~\cite{ta2022a} would reach this same significance within 2~years with a Northern area of at $40,000 \; \text{km}^2$. For building phenomenological models, a ten-fold increase in statistics would greatly improve the quality of models such as those from the combined fit~\cite{auger2017a}. For instance, the fit correctly identifies with a level of $5.4\sigma$ a model based on starburst galaxies compared to one containing only active galactic nuclei. For the benchmark area, this significance would exceed $20\sigma$ in 5~years (see ref.~\cite{gcosCompendium} for details).

The measurement of diffuse fluxes of UHE neutrinos and photons of cosmogenic origin would also provide important insights into the nature of UHECR sources~\cite{romerowolf2018a, alvesbatista2019a, heinze2019a}. An interesting example is the possibility to constrain the fraction of protons at the highest energies using neutrinos~\cite{vanvliet2019a}. GCOS could place strong constraints on this quantity~\cite{muzio2023a}, ultimately constraining the cosmological evolution of UHECR sources.

\myparagraph{Astrophysical transients.} When a transient astrophysical event emitting UHE neutral particles triggers the detector, a fast and reliable reconstruction procedure for high-quality events that satisfy specific trigger condition can be implemented and quickly sent out to partner facilities for follow-up. In addition to speed, good angular resolution (sub-degree) is also required for this purpose since other facilities have limited field of view, and the slew speeds of many telescopes can often be low (i.e., a few seconds), comparable to the characteristic time scale of some astrophysical events. For follow-ups, GCOS is exceptional, as there is no delay in pointing the apparatus to a given direction. The envisioned layout containing multiple sites ensure that a considerable fraction of the sky is continuously covered.
GCOS is planned to be distributed across multiple sites worldwide, providing coverage of the entire sky. In this case, it would be essential for GCOS to cover not only both hemispheres but also have at least four sites spread across various longitudes, to ensure that no astrophysical event would go undetected. This justifies the layouts discussed in section~\ref{sec:detector}.

\myparagraph{Dark matter.} The overwhelming body of observations supports the existence of particle dark matter (DM). Weakly interacting massive particles (WIMPs) have been popular candidates, but the searches have thus far been fruitless~\cite{bertone2018a, bertone2018b}. DM particles such as superheavy DM (SHDM) are particularly interesting. Produced in the primeval universe at the end of the inflationary phase, their decay could lead to appreciable fluxes of UHE photons compared to hadrons~\cite{anchordoqui2021a}, which could be measured with UHECR detectors~\cite{ta2019a, auger2022c, auger2023b}. In fact, these results are close to constraining many of the existing models. With an area of at least $60,000 \; \text{km}^2$, the number of events detected by GCOS would be about 20 times larger, reducing the upper limits correspondingly and consequently excluding some models.
The only requirement is good hadron/photon discrimination, which translates to $\Delta \Xmax \lesssim 30 \; \text{g} \, \text{cm}^{-2}$. A natural target for searches would then be the Galactic centre, since the DM density in this direction is expected to be higher than elsewhere. SHDM is not the only DM candidate that can be promptly studied with a facility like GCOS. Nuclearites~\cite{derujula1984a, singhsidhu2019a} and quark nuggets~\cite{bai2019a, anchordoqui2021c} would also lead to specific signatures in UHE detectors. 

\myparagraph{Particle physics.} Air-shower experiments can probe particle interactions at much higher energy scales than particle accelerators. One direct measurement that only this class of facilities can deliver are measurements of the proton-air cross section at high energies, as done by Auger~\cite{auger2012d}, TA~\cite{ta2020b}, and others. Such measurements can help distinguish among different hadronic interaction models. The expected reduction in both statistical and systematic uncertainties will enable accurate measurements at even higher energies and possibly even some model exclusions.

The reliance on hadronic interaction models is currently one of the main challenges in reconstructing EASs. This has led to different interpretations of measurements by different observatories~\cite{abbasi2015a}. Moreover, an intriguing discrepancy between the predicted and measured number of muons in air showers, according to the post-LHC hadronic interaction models, has been observed~\cite{auger2021a}. For this reason, the ability to accurately measure muons is one of the most desirable features of next-generation facilities. If the so-called `muon puzzle'~\cite{albrecht2022a} is not a result of new physics, but just a limitation inherent to the existing models,  future LHC runs with intermediate-mass nuclei, such as oxygen, are likely to provide insight and potentially resolve this issue~\cite{citron2019a, anchordoqui2022a}.

\myparagraph{Fundamental physics.} By studying the propagation of UHE particles over cosmological distances, we can search for manifestations of quantum gravity (QG) phenomena that affect the behaviour of these particles~\cite{addazi2022a}. This typically occurs as a consequence of changes in kinematical thresholds for interactions, as well as the emergence of new interactions that are normally forbidden. Depending on the particles involved, if they are unstable, their lifetimes could also be altered. Similar effects can play an important role in the development of EASs as well, impacting observables such as the muon content or its ratio with respect to the electromagnetic component of the shower.  Lorentz invariance violation (LIV) is perhaps the most extensively studied QG effect using measurements of UHE particles~\cite{saveliev2011a, anchordoqui2018c, auger2022a}. Nevertheless, other types of QG frameworks such as deformed special relativity (DSR) have also been gaining some traction~\cite{amelinocamelia2005b, lobo2022a}.

\myparagraph{Geophysics and atmospheric sciences.} With the instrumentation of GCOS in place, including fluorescence telescopes and radio antennas, it will be possible to study atmospheric transient luminous events (TLEs). They typically take place in the middle-atmosphere and are related to lightning. ELVES are a type of TLEs consisting of flashes of low-frequency radiation of characteristic size $\gtrsim 250 \; \text{km}$ and occurring in the ionosphere above thunderstorms. The Auger Collaboration recently compiled a list of nearly 1600 ELVES, making the observatory an excellent facility for this type of study~\cite{auger2020h}. Another interesting phenomenon are terrestrial gamma-ray flashes (TGFs), which are downward-beamed radiation resulting from fast cloud-to-ground discharges, linked to lightning activity, as established by observations with the surface detectors of TA~\cite{ta2018d}.

\section{The Detectors}
\label{sec:detector}

The successful realisation of the science case described in section~\ref{sec:science} requires a significantly larger and improved experiment. GCOS aims to cover an area between 40,000 and 80,000~$\text{km}^2$, utilising multiple sites to achieve full sky coverage.  With a benchmark area of $60,000 \; \text{km}^2$, we can reach the projected integrated Auger exposure (in 2030) within just one year, resulting in a ten-fold increase in statistics within a decade of operation. GCOS is expected to employ a combination of techniques, namely surface, fluorescence, and radio detectors. 

\begin{wrapfigure}{r}{0.48\textwidth}
  \centering
  \includegraphics[width=0.50\textwidth]{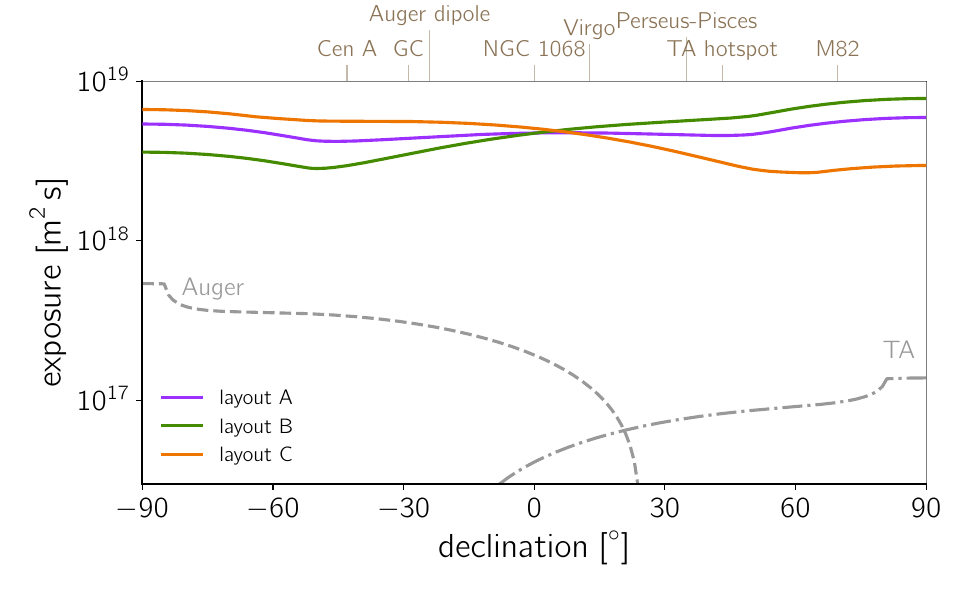}
  \caption{Exposure for layouts A, B, and C, together with those of Auger and TA. Also plotted are a few astrophysical objects and some potential anisotropy signals of interest.}
  \label{fig:exposure}
\end{wrapfigure}
We consider three representative cases for the layout, all covering a total area of $60,000 \; \text{km}^2$ distributed across sites of equal area. Sites 1 and 2 represent Auger and TA, respectively. For sites 3 and 4, their coordinates are $(\text{latitude, longitude}) = (43^\circ, 103^\circ)$ and $(-25^\circ, 137^\circ)$. Layout A is a combination of sites 1 and 2; layout B combines sites 1, 2, and 3; layout C encompasses them all. The exposure for each layout is depicted in figure~\ref{fig:exposure}, assuming that the detectors are efficient for zenith angles of up to $90^\circ$ (see section~\ref{sec:rd}). 
Note that the locations of these sites are chosen merely for illustrative purposes, to demonstrate the impact of site distribution across the globe. However, \emph{continuous} full-sky exposure is highly desirable for multimessenger studies, requiring considerations about the longitudes of the sites.

\begin{wrapfigure}{r}{0.52\textwidth}
  \centering
  \includegraphics[width=0.55\textwidth]{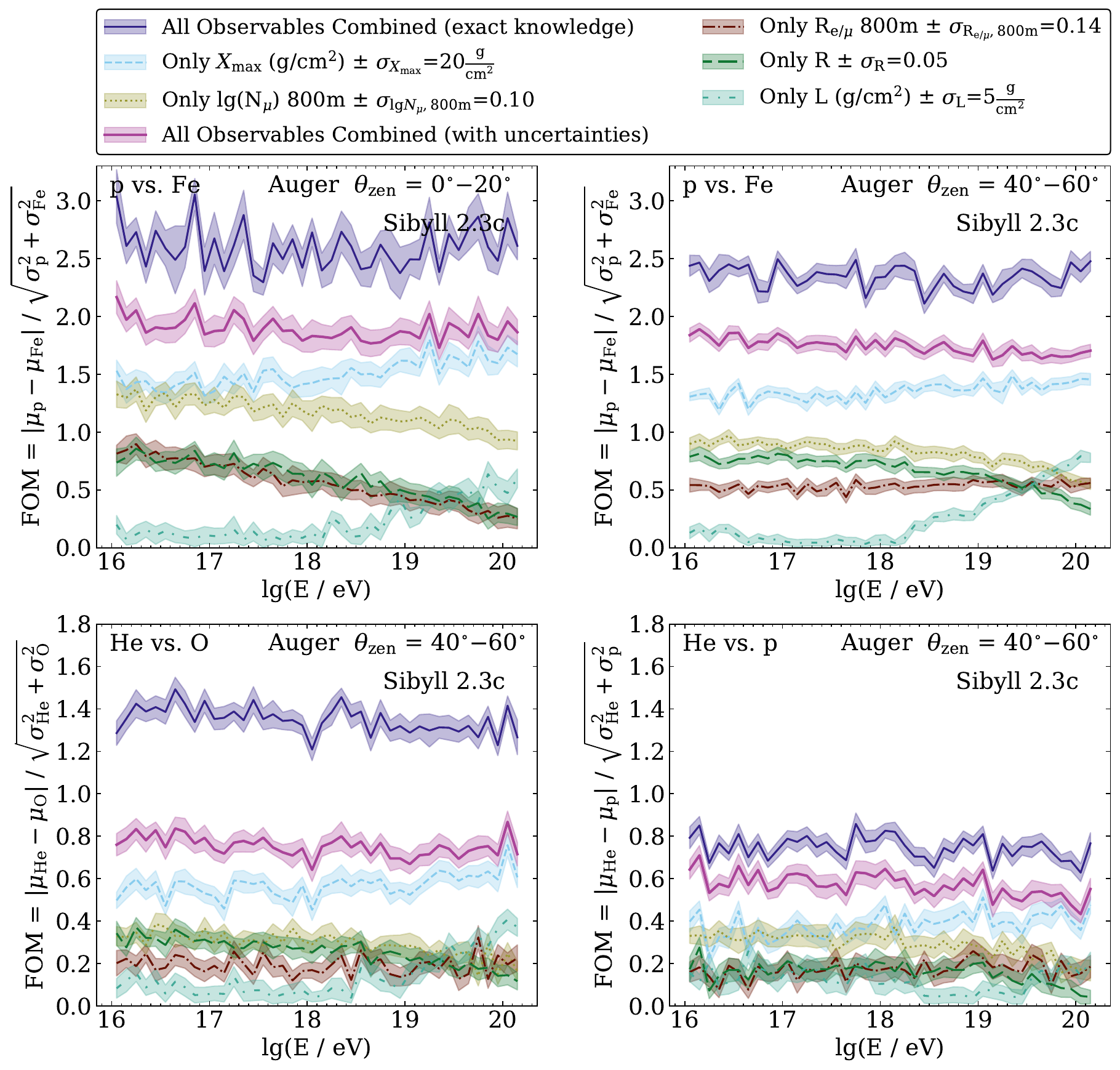}
  \caption{Figure of merit obtained from simulations for an observatory at the same location as Auger. The electron number at $\Xmax$ serves as an energy reference and was used to scale the other observables. A 10\% uncertainty is assumed for this energy reference. For details see ref.~\cite{flaggs2023a, flaggs2023b}.}
  \label{fig:merit}
\end{wrapfigure}
GCOS aims to improve mass discrimination ($\Delta \ln A < 0.8$) whilst reducing angular and energy ($< 10\%$ at the highest energies) uncertainties. The first relates to the ability of the experiment to separate individual mass groups, which depends on how well the electromagnetic and the muonic components of the shower can be separated. For $\Delta \Xmax \approx 35 \; \text{g} \, \text{cm}^{-2}$, this implies $\Delta (N_e / N_\mu) \sim 15\%$. 
Fig.~\ref{fig:merit} illustrates the capabilities of an observatory located at the site of Auger in distinguishing nuclear species using the ``figure of merit''. 
Several observables were considered, including the electromagnetic shower profile ($X_\text{max}$, $R$, $L$), the muon number at the ground within an annulus ranging from $800$ to $850 \; \text{m}$ from the shower axis, and the electron-to-muon ratio within the annulus. These results can be used to select the sites of GCOS and to help determine the optimal detectors to achieve a better mass discrimination. 

\vspace{-.2cm}
\subsection{Surface Detectors (SDs)} \label{sec:sd}

Several designs are being considered for the SDs, most of which involve the use of water-Cherenkov detectors (WCDs). 
An interesting concept being entertained are layered WCDs~\cite{letessierselvon2014b}, currently being tested at the Pierre Auger Observatory. In this design, the optical detection modules are divided into two layers at different heights relative to the bottom of the tank. Generally, the exact location of the separating sheet and the tank dimensions are optimised to obtain an approximately uniform zenithal response. 
Other possibilities include the use of scintillators or a combination of scintillators and WCDs, aiming to achieve a better separation between the electromagnetic and muonic components of the EAS~\cite{gcosCompendium}.

\begin{figure}[htb]
  \centering
  \includegraphics[width=0.32\columnwidth]{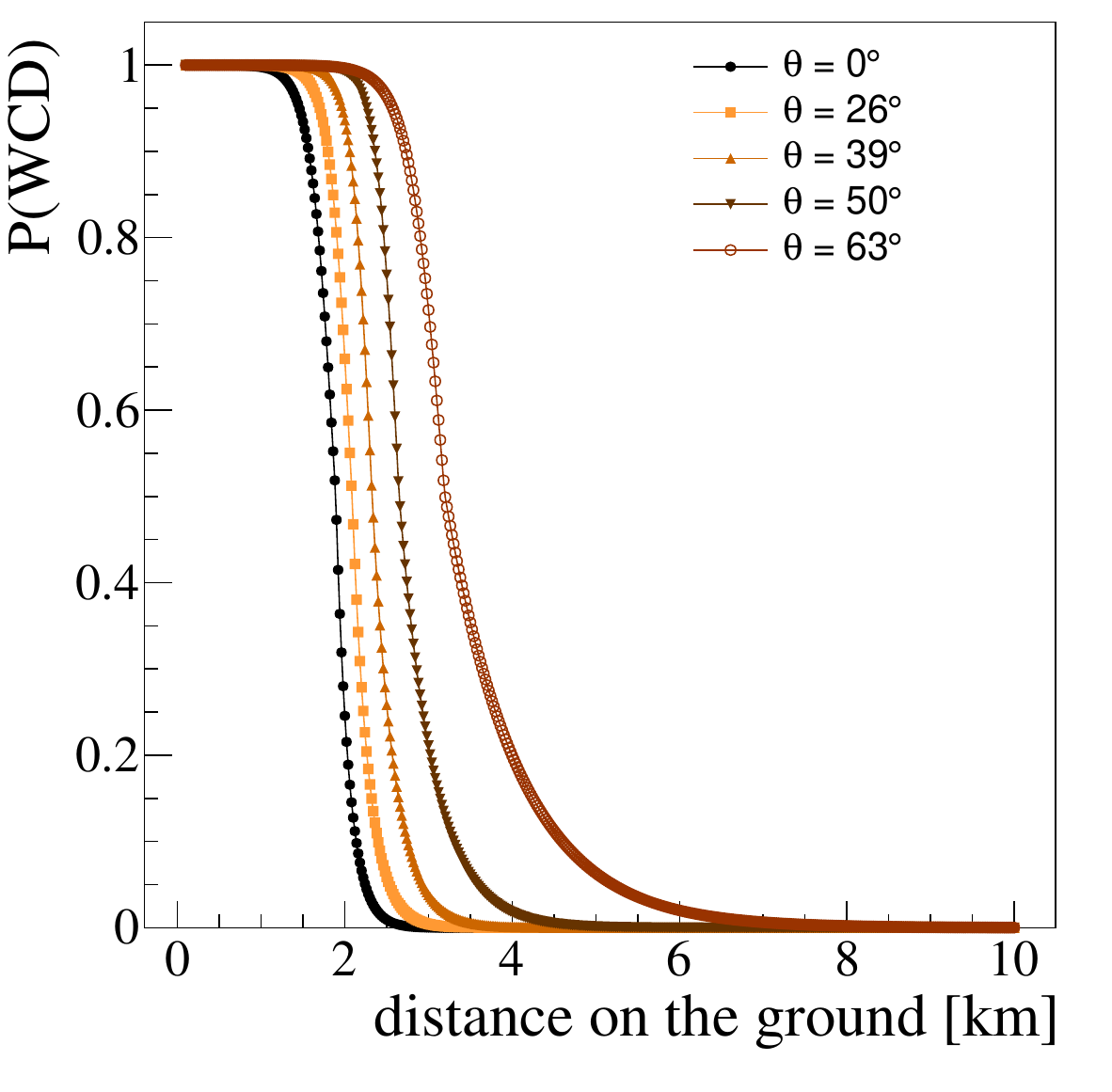}
  \includegraphics[width=0.32\columnwidth]{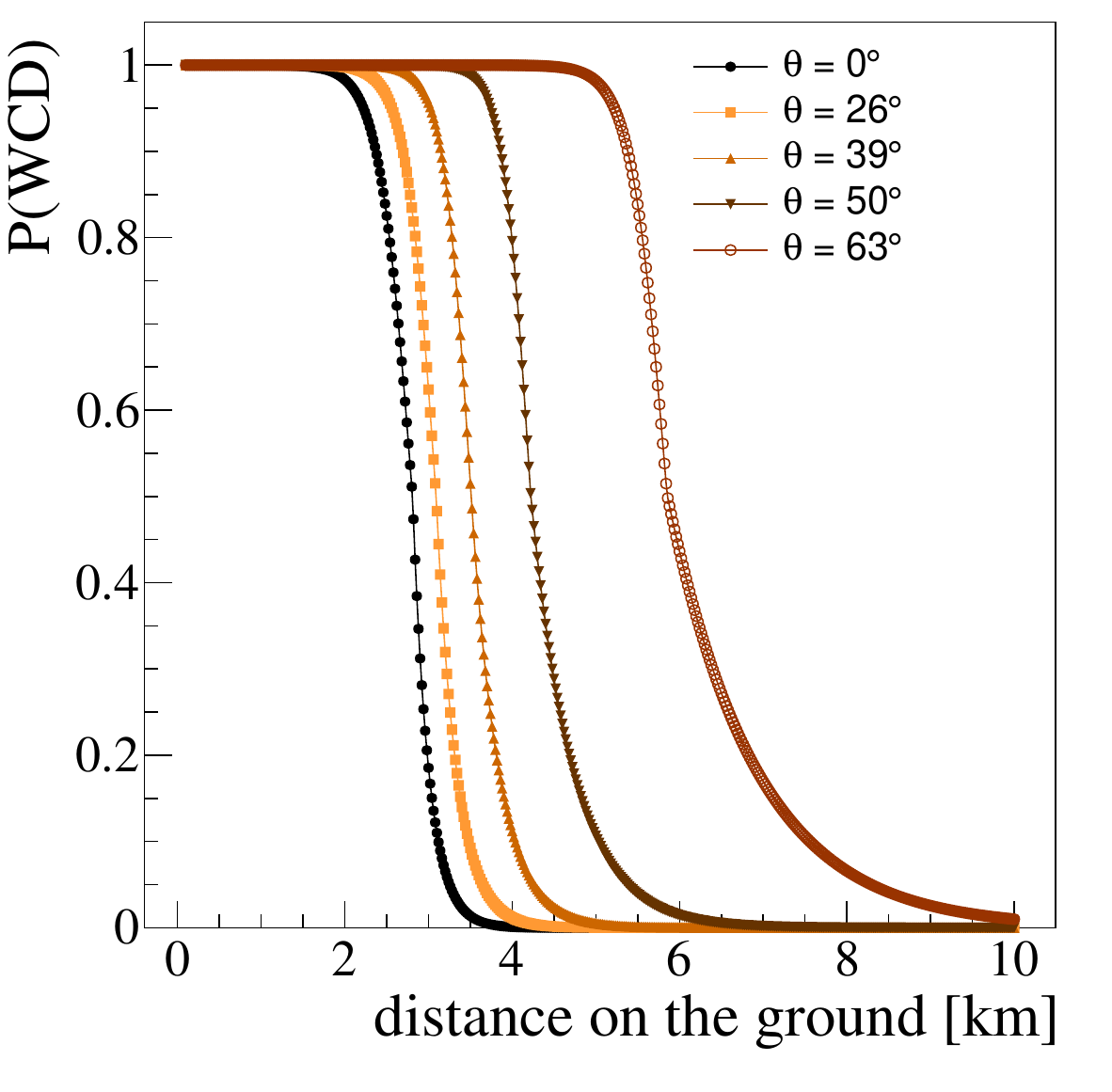}
  \includegraphics[width=0.32\columnwidth]{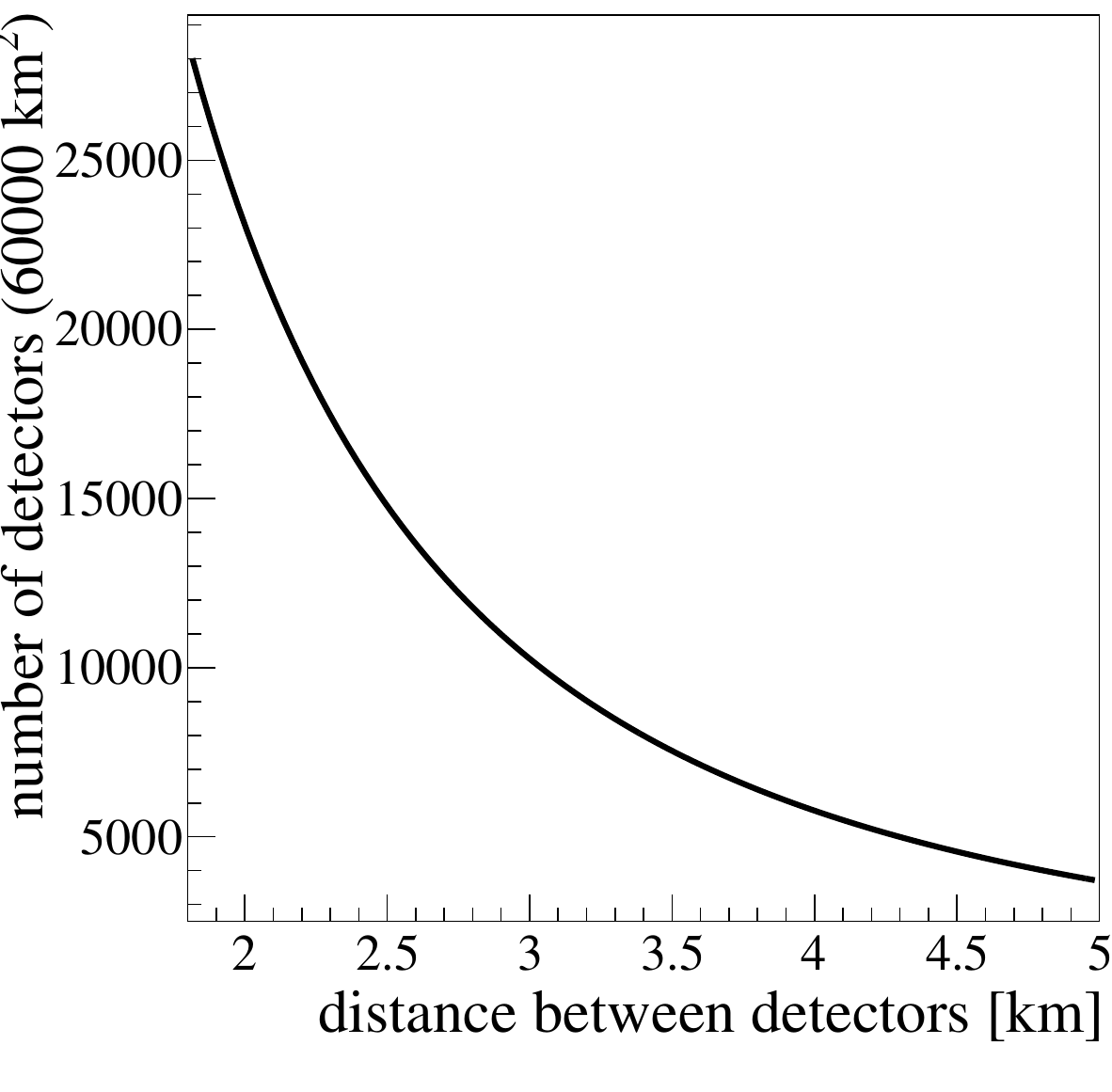}
  \caption{The left and central panels show the lateral trigger probability as a function of the distance on the ground, for various zenith angles, for energies of 10~EeV (left) and 100~EeV (right), respectively. The panel on the right shows the number of detectors required to achieve maximum efficiency at 10-30 EeV, as a function of the spacing between detectors. For details see ref.~\cite{gcosCompendium}.}
  \label{fig:layout_sd}
\end{figure}

Preliminary studies (see~\cite{gcosCompendium}) suggest that to achieve maximum efficiency at 10-30~EeV, assuming an hexagonal grid, each detector cannot be farther than approximately $2-2.5 \; \text{km}$ from each other, as show in figure~\ref{fig:layout_sd}. This would imply a total of 15k to 22k detectors. 

Photosensors in WCDs would be needed to observed the Cherenkov emission. Silicon photomultipliers (SiPMs) tend to be  more performant and energy-efficient than the traditional one, and do not require frequent maintenance. Furthermore, as the SiPM technology evolves, their cost will probably decrease considerably, making them a viable option for GCOS.

\subsection{Fluorescence Detectors (FD)} \label{sec:fd}

While SDs have a duty cycle of virtually 100\%, they fall short in fully measuring the shower, relying on models to reconstruct the primary properties. For this reason fluorescence detectors (FDs) serve as excellent complements, providing provide a calorimetric measurement of the shower and its $\Xmax$. Among the current telescope designs under consideration are the Fluorescence detector Array of Single-pixel Telescopes (FAST)~\cite{fujii2016a, malacari2020a}, the Cosmic-Ray Air Fluorescence Fresnel-lens Telescope (CRAFFT)~\cite{tameda2019a}, and a scaled-down version of the MACHETE design~\cite{cortina2016a, otte2019a} for one or a few ``cyclopes'' sites with large viewing range. 

CRAFFT has successfully detected CR showers and further tests are being performed at the TA site. Its design covers a field of view of about $24^\circ$. FAST telescopes can view an area of $30^\circ \times 30^\circ$. By combining 12 such telescopes, FAST can achieve $360^\circ$  azimuthal coverage. 
A single cyclops station, consisting of a ring with an aperture of $15 \; \text{m}^2$ and a small-pixel camera, could cover a large detection area with a field of view of $10^\circ$ in elevation $360^\circ$ in azimuth. This design is an interesting option to minimise the number of FD stations, albeit requiring additional atmospheric monitoring efforts compared to other designs.

For GCOS, interesting FD layouts could be obtained by combining cyclops and FAST or CRAFFT telescopes. The ``superman'' design would involve only three FD stations for the hybrid observations of air showers over a $40,000 \; \text{km}^2$ area. Each station would consist of one cyclops and one FAST or CRAFFT-type station, allowing for observations of far and closer showers, respectively. Another feasible option is a simple array of FDs composed of 20 FAST telescope sites covering an area of $250 \times 160 \; \text{km}^2$.

\subsection{Radio Detectors (RD)} \label{sec:rd}

Radio detection of air showers is a promising strategy with great potential~\cite{schroeder2017a}. They have been successfully used in experiments such as Auger~\cite{auger2014d, auger2016d, auger2016c, auger2018b} and LOFAR~\cite{corstanje2021a}. Like FDs, radio detectors (RDs) provide a high-precision ($\sim 10\%$) calorimetric measurement of the energy of the electromagnetic component of the shower, with the advantage of having nearly 100\% duty cycle. RDs can also deliver measurements of the depth of shower measurement with a resolution $\Delta \Xmax \sim 15 \; \text{g} \, \text{cm}^{-2}$ and angular resolutions better than $0.5^\circ$. The main difficulty is to achieve efficient self-triggering rates outside radio-quiet regions. 

Integrating radio antennas with WCDs would provide cost-effective way to improve reconstruction and enhance precision. Placing antennas on top of WCDs, for example, could improve mass sensitivity by disentangling the muonic and electromagnetic components of the showers. Ongoing antenna designs such as the prototypes being developed for GRAND~\cite{grand2020a} are interesting candidates for GCOS. A natural prototype for GCOS with demonstrated capabilities is type of antenna currently being used in Auger, placed at the top of the WCDs~\cite{hoerandel2019a}. Since the size of the radio-emission footprint strongly depends on the zenith angle of the shower~\cite{huege2016a}, a grid with spacing larger than $1.5 \; \text{km}$ would only yield access to air showers with high zenith angles. A strategy to circumvent this issue could be to deploy local clusters of radio antennas around individual WCD stations. 

Another idea to consider is interferometric shower reconstruction~\cite{schoorlemmer2021a}, already demonstrated by LOPES~\cite{lopes2021a}. This would enable the derivation of $\Xmax$ and consequently improve mass discrimination, provided that clock synchronisation between detectors is of the order of $\sim 1 \; \text{ns}$~\cite{schlueter2021a}.

\section{Outlook \& Next Steps}
\label{sec:conclusions}

The science case for a next-generation UHE experiment has been steadily advancing, building up on the findings of the Pierre Auger Observatory and the Telescope Array. The necessity to increase statistics and to reduce uncertainties, as discussed in the recent Snowmass white paper~\cite{coleman2023a}, is evident. While a concrete design has not yet been agreed upon, it is clear that GCOS will be a hybrid detector with both SDs and FDs, and RDs, covering a total area between $40,000$ and $80,000 \; \text{km}^2$ with at least two sites.
The primary goal of GCOS requires a detector optimised for the highest energies, with maximum efficiency between 10 and 30~EeV, achieving sub-degree angular resolution, as well as muon resolution better than $10\%$, and $\Delta \Xmax \lesssim 15 \; \text{g} \, \text{cm}^{-2}$.

To consolidate the scientific goals and foster innovative design ideas, three workshops were organised so far. The outcomes of these workshops have been compiled into a document serving as the foundation for future work~\cite{gcosCompendium}. Over the next three years, research and development activities will commence to construct the first prototypes. By the end of this decade, it is anticipated that at least one confirmed site will be established, where the first prototypes can be deployed.


\clearpage
\section*{Full Authors List: GCOS Collaborators}


\scriptsize
\noindent
\noindent

\noindent
M. Ahlers$^{1}$, R. Alves Batista$^{2, 3}$, P. Assis$^{4, 5}$, M. Battisti$^{6}$, J. A. Bellido$^{7}$, S. Bhatnagar $^{8}$, K. Bismark$^{9}$, T. Bister$^{10, 11}$, M. Bohacova$^{12}$, L. Caccianiga$^{13}$, R. Caruso$^{14, 15}$, W. Carvalho Jr.$^{10}$, A. Castellina$^{16, 17}$, R. Colalillo$^{18, 19}$, F. Convenga$^{20, 21}$, B. Dawson$^{7}$, I. De Mitri$^{22}$, F. de Palma$^{23}$, L. Deval$^{24}$, A. di Matteo$^{16}$, M. DuVernois$^{25, 26}$, R. Engel$^{9, 24}$, J. Eser$^{27}$, T. Fitoussi$^{24}$, B. Flaggs$^{28}$, T. Fujii$^{29}$, M. V. Garzelli$^{30}$, V. Gkika$^{31}$, P. Hamal$^{12, 32}$, B. Hariharan$^{33, 34}$, A. Haungs$^{24}$, B. Hnatyk$^{35}$, J. Hörandel$^{10}$, P. Horváth$^{32}$, M. Hrabovsky$^{32}$, T. Huege$^{24, 36}$, D. Ikeda$^{37}$, P. G. Isar$^{38}$, R. James$^{39}$, J. Johnsen$^{40}$, K.-H. Kampert$^{41}$, M. Kleifges$^{9}$, P. Koundal$^{24}$, J. Liu$^{42, 43}$, F. Longo$^{44, 45}$, D. Mandat$^{12}$, I. C. Mari\c{s}$^{46}$, J. Matthews$^{47}$, E. Mayotte$^{40}$, G. Medina-Tanco$^{48}$, K. Mulrey$^{10}$, M. S. Muzio$^{49}$, L. Nellen$^{48}$, M. Niechciol$^{50}$, S. Ogio$^{51}$, F. Oikonomou$^{52}$, M. Ostrowski$^{53}$, B. Pont$^{10}$, A. Porcelli$^{54, 55, 53}$, J. Rautenberg$^{41}$, F. Rieger$^{56}$, M. Risse$^{50}$, M. Roth$^{24}$, T. Sako$^{51}$, F. Salamida$^{20, 21}$, A. Santangelo$^{57}$, E. Santos$^{12}$, F. Sarazin$^{40}$, A. Saveliev$^{58, 59}$, M. Schimp$^{41}$, D. Schmidt$^{24}$, F. Schröder$^{28, 24}$, O. Sergijenko$^{60, 61}$, G. Sigl$^{30}$, D. Soldin$^{9, 47}$, G. Spiczak$^{62}$, Y. Tameda$^{63}$, M. Unger$^{24}$, A. van Vliet$^{64}$, S. Vorobiov$^{65}$, T. Wibig$^{66}$, B. Wundheiler$^{67}$, A. Yushkov$^{12}$, M. Zotov$^{59}$

\medskip\noindent
$^{1}$ Niels Bohr Institute; Copenhagen, Denmark\\
$^{2}$ Instituto de Física Teórica UAM-CSIC; Madrid, Spain\\
$^{3}$ Universidad Autónoma de Madrid, Departamento de Física Teórica; Madrid, Spain\\
$^{4}$ Universidade de Lisboa, Instituto Superior Técnico; Lisboa, Portugal\\
$^{5}$ Universidade de Lisboa, Laboratório de Instrumentação e Física Experimental de Partículas; Lisboa, Portugal\\
$^{6}$ Université de Paris, Astroparticule et Cosmologie; Paris, France\\
$^{7}$ The University of Adelaide, Physics Department; Adelaide, Australia\\
$^{8}$ Dayalbagh Educational Institute, Department of Physics and Computer Science; Agra, India\\
$^{9}$ Karlsruhe Institute of Technology (KIT), Institute for Experimental Particle Physics; Karlsruhe, Germany\\
$^{10}$ Radboud University, Institute for Mathematics, Astrophysics and Particle Physics; Nijmegen, Netherlands\\
$^{11}$ Nationaal Instituut voor Kernfysica en Hoge Energie Fysica (NIKHEF); Amsterdam, Netherlands\\
$^{12}$ Institute of Physics of the Czech Academy of Sciences; Prague, Czechia\\
$^{13}$ Istituto Nazionale di Fisica Nucleare (INFN), Sezione di Milano; Milan, Italy\\
$^{14}$ Department of Physics and Astronomy ``E. Majorana'', University of Catania; Catania, Italy\\
$^{15}$ Istituto Nazionale di Fisica Nucleare (INFN), Sezione di Catania; Catania, Italy\\
$^{16}$ Istituto Nazionale di Fisica Nucleare (INFN), Sezione di Torino; Turin, Italy\\
$^{17}$ Observatorio Astrofisico di Torino (INAF); Turin, Italy\\
$^{18}$ Università degli Studi di Napoli ``Federico II''; Naples, Italy\\
$^{19}$ Istituto Nazionale di Fisica Nucleare (INFN), Sezione di Napoli; Naples, Italy\\
$^{20}$ Università dell'Aquila, Dipartimento di Scienze Fisiche e Chimiche; L'Aquila, Italy\\
$^{21}$ Istituto Nazionale di Fisica Nucleare (INFN), Laboratori Nazionali del Gran Sasso; Assergi (L'Aquila), Italy\\
$^{22}$ Gran Sasso Science Institute; L'Aquila, Italy\\
$^{23}$ Università del Salento, Dipartimento di Matematica e Fisica “E. De Giorgi”; Lecce, Italy\\
$^{24}$ Karlsruhe Institute of Technology (KIT), Institute for Astroparticle Physics; Karlsruhe, Germany\\
$^{25}$ University of Wisconsin-Madison, Department of Physics; Madison, WI, USA\\
$^{26}$ University of Wisconsin-Madison, Wisconsin IceCube Particle Astrophysics Center; Madison, WI, USA\\
$^{27}$ The University of Chicago, Department of Astronomy and Astrophysics; Chicago, IL, USA\\
$^{28}$ University of Delaware, Bartol Research Institute; Newark, DE, USA\\
$^{29}$ Osaka Metropolitan University, Graduate School of Science; Osaka, Japan\\
$^{30}$ Universität Hamburg, II. Institut für Theoretische Physik; Hamburg, Germany\\
$^{31}$ Institute for Basic Science,  Center for Axion and Precision Physics Research; Daejeon, South Korea\\
$^{32}$ Palacký University Olomouc; Olomouc, Czechia\\
$^{33}$ Tata Institute of Fundamental Research; Mumbai, India\\
$^{34}$ The GRAPES-3 Experiment, Cosmic Ray Laboratory; Ooty, India\\
$^{35}$ Taras Shevchenko National University of Kyiv, Astronomical Observatory; Kyiv, Ukraine\\
$^{36}$ Vrije Universiteit Brussel, Astrophysical Institute; Brussels, Belgium\\
$^{37}$ Kanagawa University; Kanagawa, Japan\\
$^{38}$ Institute of Space Science; Magurele, Romania\\
$^{39}$ Case Western Reserve University ; Cleveland, OH, USA\\
$^{40}$ Colorado School of Mines; Golden, CO, USA\\
$^{41}$ Bergische Universität Wuppertal, Department of Physics; Wuppertal, Germany\\
$^{42}$ Key Laboratory of Particle Astrophysics, Institute of High Energy Physics, Chinese Academy of Sciences; Beijing, China\\
$^{43}$ TIANFU Cosmic Ray Research Center; Chengdu, China\\
$^{44}$ Università di Trieste, Dipartimento di Fisica; Trieste, Italy\\
$^{45}$ Istituto Nazionale di Fisica Nucleare (INFN), Sezione di Trieste; Trieste, Italy\\
$^{46}$ Université Libre de Bruxelles; Brussels, Belgium\\
$^{47}$ University of Utah, Department of Physics and Astronomy; Salt Lake City, UT, USA\\
$^{48}$ Universidad Nacional Autónoma de México, Instituto de Ciencias Nucleares; Mexico City, Mexico\\
$^{49}$ Pennsylvania State University, Department of Physics; State College, PA, USA\\
$^{50}$ University of Siegen, Center for Particle Physics; Siegen, Germany\\
$^{51}$ University of Tokyo, Institute for Cosmic Ray Research; Chiba, Japan\\
$^{52}$ Norwegian University of Science and Technology, Institutt for fysikk; Trondheim, Norway\\
$^{53}$ Jagiellonian University, Astronomical Observatory; Krakow, Poland\\
$^{54}$ Centro Ricerche Enrico Fermi (CREF); Rome, Italy\\
$^{55}$ Istituto Nazionale di Fisica Nucleare (INFN), Laboratori Nazionali di Frascati; Frascati, Italy\\
$^{56}$ Universität Heidelberg, Institut für Theoretische Physik; Heidelberg, Germany\\
$^{57}$ Universität Tübingen, Institut für Astronomie und Astrophysik; Tübingen, Germany\\
$^{58}$ Immanuel Kant Baltic Federal University; Kaliningrad, Russia\\
$^{59}$ Lomonosov Moscow State University; Moscow, Russia\\
$^{60}$ Main Astronomical Observatory of the National Academy of Sciences of Ukraine; Kyiv, Ukraine\\
$^{61}$ AGH University of Science and Technology; Krakow, Poland\\
$^{62}$ University of Wisconsin River Falls; River Falls, WI, USA\\
$^{63}$ Osaka Electro-Communication University, Graduate School of Engineering; Osaka, Japan\\
$^{64}$ Khalifa University, Department of Physics; Abu Dhabi, United Arab Emirates\\
$^{65}$ University of Nova Gorica, Center for Astrophysics and Cosmology; Nova Gorica, Slovenia\\
$^{66}$ University of Lodz, Faculty of Physics and Applied Informatics; Lodz, Poland\\
$^{67}$ Instituto de Tecnologías en Detección y Astropartículas (CNEA, CONICET, UNSAM); Buenos Aires, Argentina\\

\section*{Acknowledgements}

\noindent
RAB is funded by: ``la Caixa'' Foundation (ID 100010434) and the European Union's Horizon~2020 research and innovation program under the Marie Skłodowska-Curie grant agreement No~847648, fellowship code LCF/BQ/PI21/11830030.

\end{document}